# From $J_{eff}$=1/2 insulator to p-wave superconductor in single-crystal $Sr_2Ir_{1-x}Ru_xO_4$ ($0 \leq x \leq 1$)

S. J. Yuan[1], S. Aswartham[1], J. Terzic[1], H. Zheng[1], H. D. Zhao[1], P. Schlottmann[2] and G. Cao[1]

[1]*Center for Advanced Materials, Department of Physics and Astronomy, University of Kentucky, Lexington, Kentucky 40506, USA*
[2]*Physics Department, Florida State University, Tallahassee, FL 32306, USA*

## ABSTRACT

$Sr_2IrO_4$ is a magnetic insulator assisted by strong spin-orbit coupling (SOC) whereas the $Sr_2RuO_4$ is a p-wave superconductor. The contrasting ground states have been shown to result from the critical role of the strong SOC in the iridate. Our investigation of structural, transport, and magnetic properties reveals that substituting 4d $Ru^{4+}$ ($4d^4$) ions for 5d $Ir^{4+}$($5d^5$) ions in $Sr_2IrO_4$ directly adds holes to the $t_{2g}$ bands, reduces the SOC and thus rebalances the competing energies in single-crystal $Sr_2Ir_{1-x}Ru_xO_4$. A profound effect of Ru doping driving a rich phase diagram is a structural phase transition from a distorted $I4_1/acd$ to a more ideal $I4/mmm$ tetragonal structure near x=0.50 that accompanies a phase transition from an antiferromagnetic-insulating state to a paramagnetic-metal state. We also make a comparison drawn with Rh doped $Sr_2IrO_4$, highlighting important similarities and differences.

PACS numbers: 71.70.Ej, 75.30.Gw, 71.30.+h

## I. INTRODUCTION

The 5d-electron based iridates have continuously attracted considerable interest as they display unusual properties primarily resulting from a delicate interplay between strong spin-orbit coupling (SOC) and other competing energies such as Coulomb interactions, non-cubic crystalline electric fields, and Hund's rule coupling [1-3]. The $J_{eff}$=1/2 insulating state is a manifestation of physics driven by such a new hierarchy of energies [1,2,4].

Among all the iridates studied, the single-layered $Sr_2IrO_4$ has been subjected to the most extensive investigations due to its $J_{eff}$ = 1/2 insulating ground state, and similarities of its crystallographic, electronic, and magnetic structures to those of the undoped high-$T_C$ cuprate $La_2CuO_4$. However, $IrO_6$ octahedra in $Sr_2IrO_4$ rotate about the c-axis by about $12^o$; this distinct structural feature, which is absent in $La_2CuO_4$, critically affects the ground state of the iridate. $Sr_2IrO_4$ undergoes an antiferromagnetic (AFM) ordering at $T_N$ = 240 K, and exhibits a canted magnetic structure that rigidly tracks the staggered rotation of the $IrO_6$ octahedra in $Sr_2IrO_4$ [5-8].

It is useful to first compare $Sr_2IrO_4$ with its isostructural 4d-based counterparts, $Sr_2RhO_4$ and $Sr_2RuO_4$. Their underlying structural and physical properties are listed in Table 1 for contrast and comparison. Both $Sr_2IrO_4$ and $Sr_2RhO_4$ crystallize in a reduced tetragonal structure with space-group $I4_1/acd$ due to a rotation of the $IrO_6$ or $RhO_6$ octahedra about the c-axis by $\sim 12^\circ$ or $\sim 9.7^\circ$, respectively, resulting in an expanded unit cell by $\sqrt{2}\times\sqrt{2} \times 2$, as compared to the undistorted cell [9,10]. Despite the structural similarity, $Sr_2RhO_4$ is a paramagnetic (PM), correlated metal, sharply contrasting the magnetic insulator $Sr_2IrO_4$ [5,6,9,11,12], owed chiefly to the weaker SOC ($\sim$0.15 eV), compared with SOC ($\sim$0.4 eV) for $Sr_2IrO_4$, that renders a smaller splitting between the $J_{eff}$ = 1/2 and $J_{eff}$ = 3/2 bands [1,13]. On the other hand, $Sr_2RuO_4$ adopts an ideal tetragonal structure without the rotation of $RuO_6$ octahedra and supports a *p*-wave superconducting state [14]. Indeed, the impact of the SOC strongly depends on the detailed band structure near the Fermi surface $E_F$, the Coulomb interactions, and the lattice distortions [15-18], and this in part explains differences between the superconducting $Sr_2RuO_4$ and metallic $Sr_2RhO_4$ that is very close to the borderline of a metal-insulator transition.

Table 1 **Comparison for** $Sr_2IrO_4$, $Sr_2RhO_4$, and $Sr_2RuO_4$ [3]

| Compound | Space group | SOC (eV) | Exemplary Phenomena |
|---|---|---|---|
| $Sr_2IrO_4$ | $I4_1/acd$ | ~ 0.40 | Antiferromagnet / $J_{eff}$=1/2 insulator |
| $Sr_2RhO_4$ | $I4_1/acd$ | ~ 0.16 | Paramagnet/metal |
| $Sr_2RuO_4$ | $I4/mmm$ | ~ 0.15 | Paramagnet/p-wave superconductor at low T |

In our previous work, we tuned the ground state by substituting Rh for Ir in $Sr_2IrO_4$, in an attempt to reduce the SOC [13]. This chemical substitution generates a rich phase diagram for $Sr_2Ir_{1-x}Rh_xO_4$ ($0 \leq x \leq 1$), where a robust metallic state is not fully established until x approaches 1 due in part to a variation of the valence state of Rh with x [13,19,20]. As a natural extension of this study, we have extended our investigation to Ru doped $Sr_2IrO_4$ or $Sr_2Ir_{1-x}Ru_xO_4$.

In this paper, we report a thorough investigation of structural, transport, and magnetic properties of single-crystal $Sr_2Ir_{1-x}Ru_xO_4$ with $0 \leq x \leq 1$. Ru doping induces a structural phase transition from a distorted tetragonal structure with $I4_1/acd$ to a more ideal one with $I4/mmm$ near x=0.50. It is this structural change that marks a concurrent phase transition from the AFM insulating state (x<0.50) to a Ru-doping induced PM metallic state (x>0.50). We also make a comparison between $Sr_2Ir_{1-x}Ru_xO_4$ and $Sr_2Ir_{1-x}Rh_xO_4$, highlighting important similarities and differences.

## II. EXPERIMENT

The single crystals $Sr_2Ir_{1-x}Ru_xO_4$ were grown from off stoichiometric quantities of $SrCl_2$, $SrCO_3$, $IrO_2$, and $RuO_2$ using self-flux techniques. Similar technical details are described elsewhere [4,6,21,22]. The structures of the crystals were determined using a Nonius Kappa CCD x-ray diffractometer at 90 K. Structures were refined by full-matrix least squares using the SHELX-97 programs [23]. All structures affected by absorption and extinction were corrected by comparison of symmetry-equivalent reflections using the program SADABS [23]. It needs to be emphasized that the single crystals are of high quality and there is no indication of any mixed phases in all doped single crystals studied. The presence of any mixed phases or inhomogeneity in the single crystals would not allow any converging structural refinements. The standard deviations of all lattice parameters and interatomic distances are smaller than 0.1%. Chemical compositions of the single crystals were estimated using both single-crystal x-ray diffraction and energy dispersive X-ray analysis (Hitachi/Oxford 3000). Magnetization and electrical resistivity were measured using either a Quantum Design MPMS-7 SQUID Magnetometer and/or Physical Property Measurement System with 14-T field capability.

## III. RESULTS AND DISCUSSION

The Ru ion tends to be tetravalent $Ru^{4+}$ in perovskite ruthenates [3]. Substituting $Ru^{4+}$ ($4d^4$) for $Ir^{4+}$ ($5d^5$) in $Sr_2IrO_4$ changes the crystal structure and adds holes to the $t_{2g}$ bands. We first examine changes of the crystal structure in $Sr_2Ir_{1-x}Ru_xO_4$. $Sr_2IrO_4$ crystallizes in a distorted tetragonal structure with reduced space-group symmetry $I4_1/acd$ due to a rotation of the $IrO_6$ octahedra about the *c* axis by ~12° with the lattice parameters a=b=5.4773(8) Å and c=25.76(5) Å at T=90 K. This rotation corresponds to a distorted in-plane Ir-O1-Ir bond angle θ (=156.474° at T=90 K). In sharp contrast, $Sr_2RuO_4$ crystallizes in the ideal $K_2NiF_4$ structure with space group *I*4/*mmm* featuring 180° Ru-O1-Ru bonds in the basal plane or no rotation of $RuO_6$ octahedra [10]. With increasing x, Ru doping initially weakens and eventually eliminates the structural distortions with a decrease in the lattice parameters *a*- and *c*- axis and the ratio of *c*/*a*, as shown in Fig. 1. More importantly, a structural transition from $I4_1/acd$ to *I*4/*mmm* occurs near x=0.50. The Ir/Ru-O1-Ir/Ru bond angle θ, reflecting the rotation of the octahedra about the *c*-axis, increases with x and becomes 180° abruptly near x=0.50, the structural transition (see Fig. 2(a)). The in-plane bond length Ir/Ru-O1 shortens correspondingly with a sudden shortening at the structural transition as well; it then levels off with further increasing x, as shown in Fig. 2(b). On the other hand, the Ir/Ru-O2 bond length, which is more closely associated with the lattice parameter c-axis, initially decreases with x, and then shows a sudden increase at x=0.50 before decreasing again with further increasing x (see Fig. 2(c)). For contrast and comparison, we also illustrate the lattice parameters of $Sr_2Ir_{1-}xRh_xO_4$ (see Fig. 2(d)-(f)). Apparently, all the bond angle and bond lengths for Rh doped samples show only slight changes with increasing x, sharply contrasting with those in the Ru doped $Sr_2IrO_4$.

The electrical resistivity ρ(T) of $Sr_2Ir_{1-x}Ru_xO_4$ for the *a* and *c* axes drastically reduces by nearly five orders of magnitude at low temperatures as x is increased from x=0 to 0.17, and a metallic state is induced at x=0.49 [see Figs. 3(a) and 3(b)]. For x≥0.49, there is an upturn at low T, in the *a*-axis resistivity $\rho_a$(T). The temperature of the minimum is denoted with T*, which decreases with x. A metal state is only fully realized at x=0.92. This behavior is similar to that observed in $Sr_3(Ir_{1-x}Ru_x)_2O_7$; it is attributed to a robust Mott gap that blocks the charge transfer of doped holes [24]. The c-axis resistivity $\rho_c$ exhibits a different temperature-dependence and larger magnitude, particularly for heavier Ru doped $Sr_2IrO_4$. The increased anisotropy in ρ(T) suggests a two-dimensional nature of the electronic structure and is qualitatively consistent with the changes in the in-plane and out-of-plane Ir/Ru-O bond lengths (Figs. 2b and 2c). It is remarkable that the resistivity exhibits no discernible effect due to disorder in $Sr_2Ir_{1-x}Ru_xO_4$. In contrast for Rh substitution the system always remains in the proximity to the insulting state. Each Ru atom adds one hole, which gives rise to a higher density of states near $E_F$; more importantly, Ru doping drives a structural phase transition to an ideal tetragonal structure with no octahedral distortion, thus enhances the electron hopping, and supports a more robust metallic state in $Sr_2Ir_{1-x}Ru_xO_4$ when x approaches 1. Under these circumstances disorder in the alloy plays a less relevant role, in contrast to the situation in Rh doped $Sr_2IrO_4$ in which Anderson localization dominates a wide range of Rh doping [13].

The temperature-dependent magnetization M(T) data for representative compositions of single crystals $Sr_2Ir_{1-x}Ru_xO_4$ are presented in Fig. 4. There is a kink in M(T) for x=0 at 100 K that is attributed to a possible rearrangement of the magnetic order and is closely associated with magnetoresistivity [25], magnetoelectric effect [4] and unusual muon responses [26]. Ru doping suppresses the AFM transition $T_N$ from 240 K at x = 0 to zero at x = 0.49. It needs to be pointed out that the AFM transition $T_N$ for 0.40 < x <0.49 becomes less well-defined, however, a close examination indicates that the $T_N$ is not completely suppressed to zero until x=0.49. Nevertheless, it is reasonably close to the classical (*i.e.* spin-only) two-dimensional site percolation threshold of x = 0.41 [27]. It is also noted that the AFM state vanishes at x=0.16 in Rh doped $Sr_2IrO_4$ or $Sr_2Ir_{1-x}Rh_xO_4$ [13]. The rapid suppression of the AFM state is attributed to a varying valence state of Ir and Rh ions and a change in the relative strength of SOC, tetragonal electric field effects and Hund's rule coupling, which competes with the SOC and prevents the $J_{eff}$=1/2 state [13,19].

We analyzed the magnetic data using the Curie-Weiss law, $\chi = \chi_0 + C/(T - \theta_{CW})$ (where $\chi_0$ is a temperature-independent constant, $\theta_{CW}$ the Curie-Weiss temperature, and $C$ the Curie constant) and then used $\chi_0$ to obtain $\Delta\chi = \chi - \chi_0 = C/(T - \theta_{CW})$ and $\Delta\chi^{-1}$ vs $T$, as shown in Fig. 5(a). Here, $C=\frac{N_A}{3k_B}\mu_{eff}^2$, with $N_A$ being Avogadro's number and $k_B$ the Boltzmann constant. The effective magnetic moment $\mu_{eff}$ per formula unit is then derived from C, as shown in Fig. 5(b). Note that the temperature range for the fit depends on x, but a high temperature interval is used in every case. $\mu_{eff}$ remains essentially unchanged initially and then increases rapidly when x > 0.49, peaking at x = 0.58 before decreasing with further increasing x. The peak happens in the doping range where the structural phase transition takes place (see Fig. 2(a)-(c)). The Ru doping dependence of $\mu_{eff}$ is qualitatively consistent with the results in an earlier study on polycrystalline samples [28]. The Curie-Weiss temperature $\theta_{CW}$ tracks $T_N$ for $0 \leq x \leq 0.49$, and then changes its sign from positive to negative as $x$ increases further, as shown in Fig. 5(c). It is remarkable that the abrupt change in $\theta_{CW}$ also occurs in the range of the structural phase transition, echoing the sudden jump of $\mu_{eff}$. $\theta_{CW}$ was obtained from a high T fit and it is positive (ferromagnetic exchange) in the antiferromagnetic region, where $T_N > 0$. Note also that $\theta_{CW} = -126$ K for x=0.58 where no long-range order exists. Since $\theta_{CW}$ measures the strength of the magnetic interaction, such a large absolute value of $\theta_{CW}$ in a system without magnetic ordering implies a strong magnetic frustration, which may primarily result from a competition between the AFM (Ir 5d-electrons) and ferromagnetic (Ru 4d-electrons) coupling.

Ru doping affects the magnetic anisotropy as well. The c-axis magnetization $M_c$ becomes stronger than the a-axis magnetization $M_a$, especially at low temperatures, with increasing x (see Fig. 6 as well as Fig. 4). This behavior is absent in Rh doped $Sr_2IrO_4$ but is observed in $Ca_2Ru_{1-x}Ir_xO_4$ due to the strong interaction between Ru 4d- and Ir 5d-electrons [29]. For x=0, $M_a$ is larger than $M_c$ because the magnetic moment lies within the basal plane [7]. Upon Ru doping, $M_c$ becomes larger than $M_a$ at low temperatures initially and then throughout the entire temperature range measured for x ≥ 0.58 (see Fig. 6). This change suggests a spin flop from the basal-plane to the c-axis due to Ru doping. Interestingly, Rh doping (up to x=0.12) rearranges the in-plane magnetic configuration without any c-axis magnetic component [13].

The above evolution of the transport and magnetic properties closely follow the changes in the lattice properties. As illustrated in Fig. 2(a)&(b), Ru doping results in an increase in the Ir/Ru-O1-Ir/Ru bond angle and a decrease in the in-plane Ir/Ru-O1 bond length, which inevitably enhance the d-orbital overlap or electron hopping. These lattice changes along with added holes and reduced SOC explain the drastic decrease in the electrical resistivity (Fig. 3) and the vanishing AFM state.

A phase diagram for $Sr_2Ir_{1-x}Ru_xO_4$ generated based on the data presented above summarizes the central findings of this study, as shown in Fig. 7. The most prominent feature of the phase diagram is the structural phase transition from a distorted $I4_1/acd$ to a more ideal $I4/mmm$ tetragonal structure near x=0.50; this structural phase transition accompanies a magnetic transition from the canted-antiferromagnetic-insulating (CAF-I) to paramagnetic-metal (PM-M) ground state. All results indicate that the $Ru^{4+}(4d^4)$ substituting $Ir^{4+}(5d^5)$ adds holes into the $t_{2g}$ bands and reduces SOC but it is the lattice degrees of freedom that primarily drive the rich phase diagram. Remarkably, this phase diagram contrasts with that of Rh doped $Sr_2IrO_4$ [Fig. 5 in Ref. 13] in which the AFM state vanishes more rapidly (at 16% Rh doping) but the insulating state is much more resilient to Rh doping in part because of the rotation of $RhO_6$ octahedra in $Sr_2RhO_4$ that leads to a band folding and narrowing, giving rise to nearly degenerate states close to the Fermi level [17] and because of the varying valence state of both Rh and Ir that causes the Anderson localization [13,19,20].

## ACKNOWLEDGMENTS

This work was supported by the National Science Foundation via Grant No. DMR-1265162 and by Department of Energy (BES) through grant No. DE-FG02-98ER45707 (PS).

*Corresponding authors: shujuan.yuan@uky.edu; cao@uky.edu

## REFERRENCES

## Figure Captions

**FIG. 1.** Ru concentration $x$-dependence at $T$ = 90 K of the lattice parameters of the $a$ axis (a), the $c$ axis (b) and the $c/a$ ratio (c). Inset: representative single-crystal Bragg diffraction peaks for the [001] direction; note the highly ordered crystal structure of $Sr_2Ir_{1-x}Ru_xO_4$. The shaded area indicates the region where the structural phase transition occurs.

**FIG. 2.** On the left panel, the Ru concentration $x$-dependence at T = 90 K of (a) the Ir/Ru-O1-Ru/Ir bond angle θ, (b) the in-plane Ir/Ru-O1 bond length and (c) the out-of-plane Ir/Ru-O2 bond length. The shaded area indicates the region where the structural phase transition occurs. For comparison, the right panel shows the Rh concentration $x$-dependence of the Ir/Rh-O1-Ir/Rh bond angle θ, (b) the in-plane Ir/Rh-O1 bond length and (c) the Ir/Rh-O2 bond length. The data for Rh doping is obtained from the crystals used in Ref. 13. The insets show the definition of the bond angle Ir/Ru-O1-Ir/Ru, and the bond lengths Ir/Ru-O1 and Ir/Ru-O2.

**FIG. 3.** The temperature dependence of the resistivity $\rho(T)$ in the ***ab***-plane (a)&(b) and along the ***c***-axis (c)&(d) for representative compositions x = 0, 0.17, 0.36, 0.49, 0.58, 0.65, 0.74, and 0.92. The arrows indicate the minimum of $\rho_a(T)$ defining T*.

**FIG. 4.** The temperature dependence at $\mu_0 H = 0.1$ T of the magnetization (a) $\boldsymbol{M}_a$ and $\boldsymbol{M}_c$ for x=0, (b) $\boldsymbol{M}_a$ and (c) $\boldsymbol{M}_c$ for the representative compositions x = 0, 0.17, 0.25, 0.40, 0.58, 0.74, and 0.92.

**FIG. 5.** (a) The temperature dependence of magnetic susceptibility $\Delta\chi^{-1}$ for the representative compositions x = 0.17, 0.40, 0.58, and 0.74. The Ru concentration $x$-dependence of (b) the magnetic effective moment $\mu_{eff}$, and (c) $T_N$ and $\theta_{CW}$. Note the varying temperature intervals for the fit.

**FIG. 6.** The temperature dependence at $\mu_0 H = 0.1$ T of the magnetization $\boldsymbol{M}_a$ and $\boldsymbol{M}_c$ for representative compositions (a) x=0.17, (b) x=0.25, (c) x=0.4, and (d) 0.58. The magnetization was measured after field cooling at $\mu_0 H = 0.1$ T.

**FIG. 7.** The phase diagram for $Sr_2Ir_{1-x}Ru_xO_4$ generated based on the data presented above. Note that CAF-I denotes the canted antiferromagnetic insulating phase, PM-I denotes the paramagnetic insulating phase, PM-M indicates the paramagnetic metallic regime.

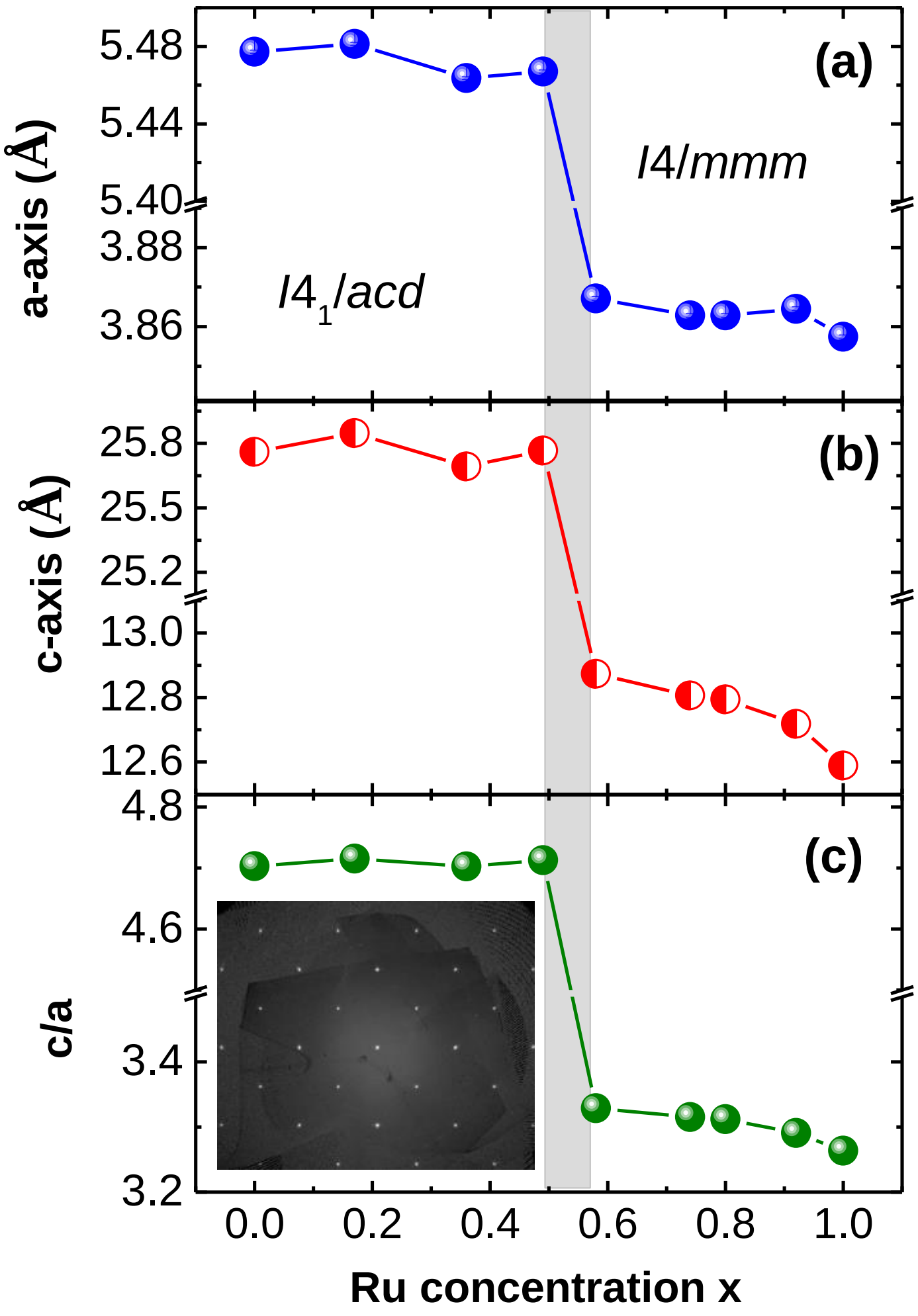

Fig. 1

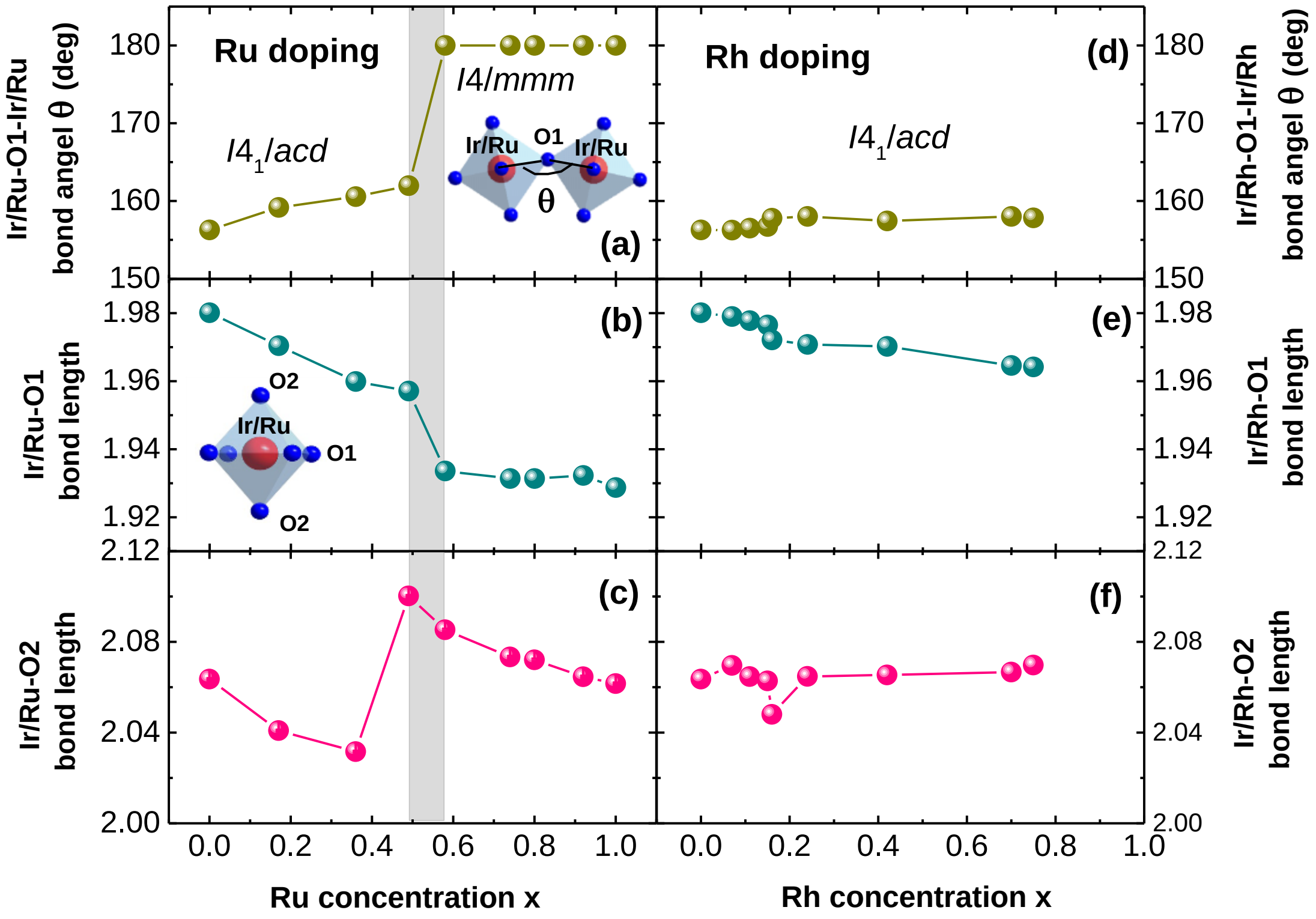

Fig. 2

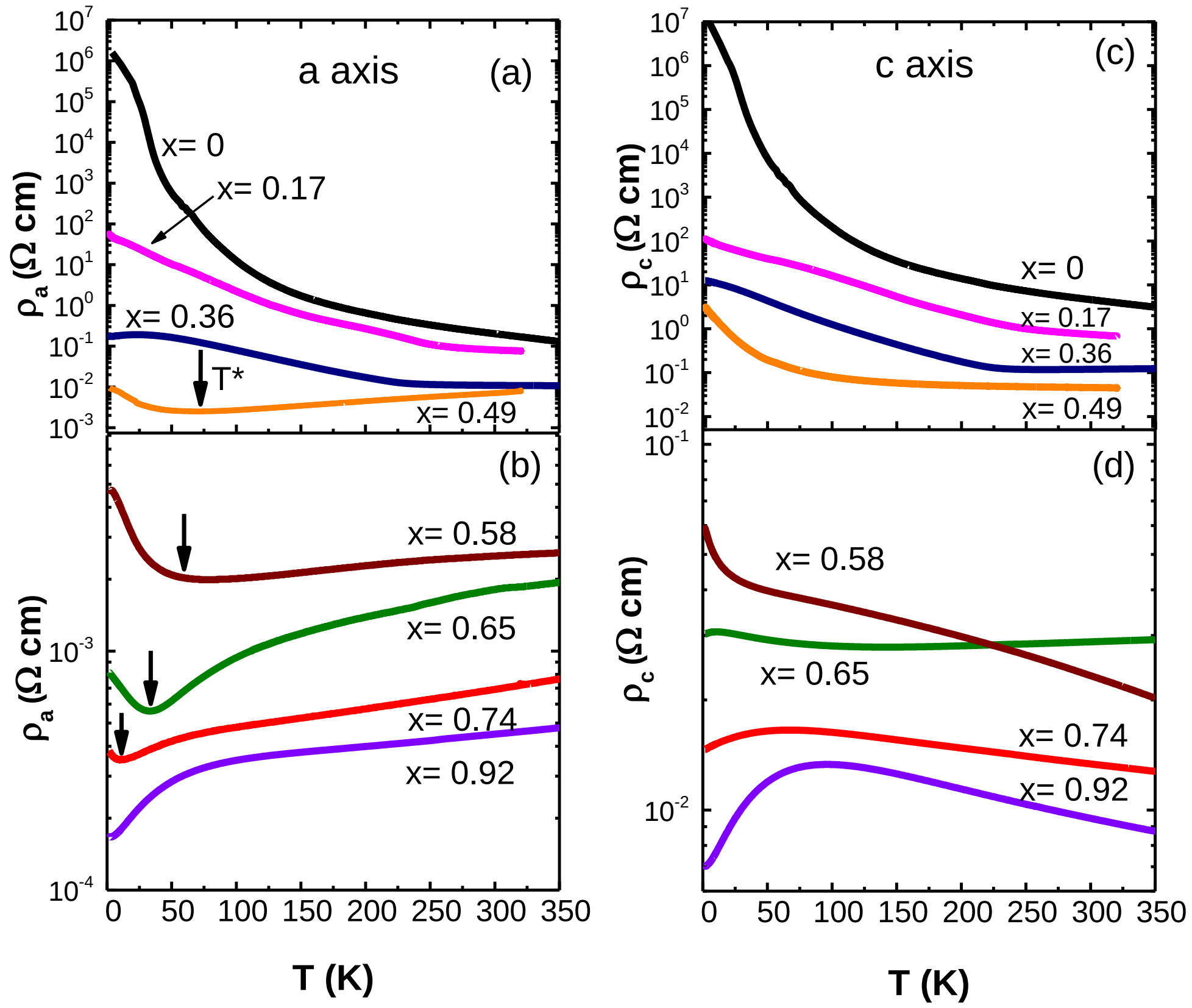

Fig. 3

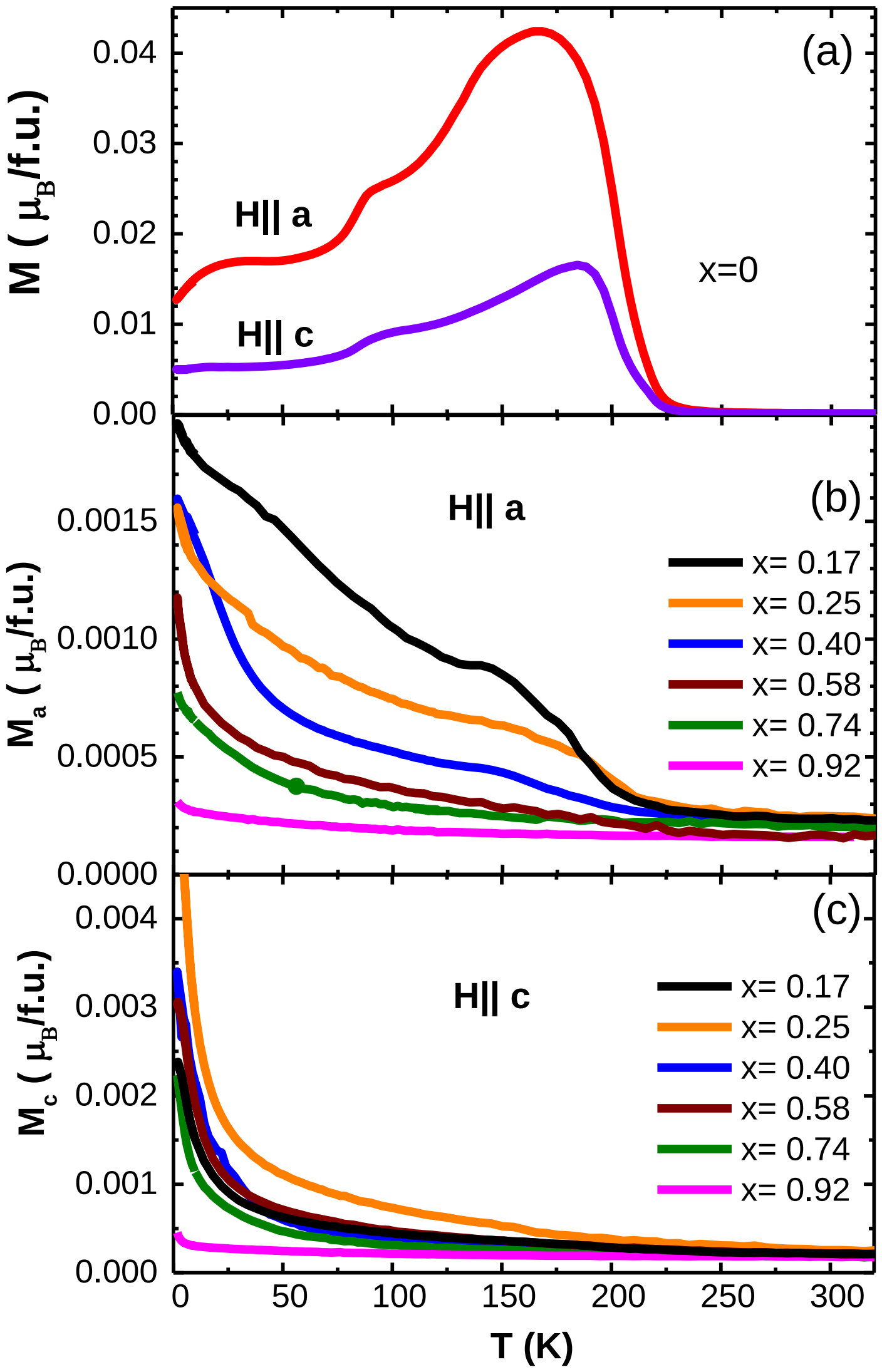

Fig. 4

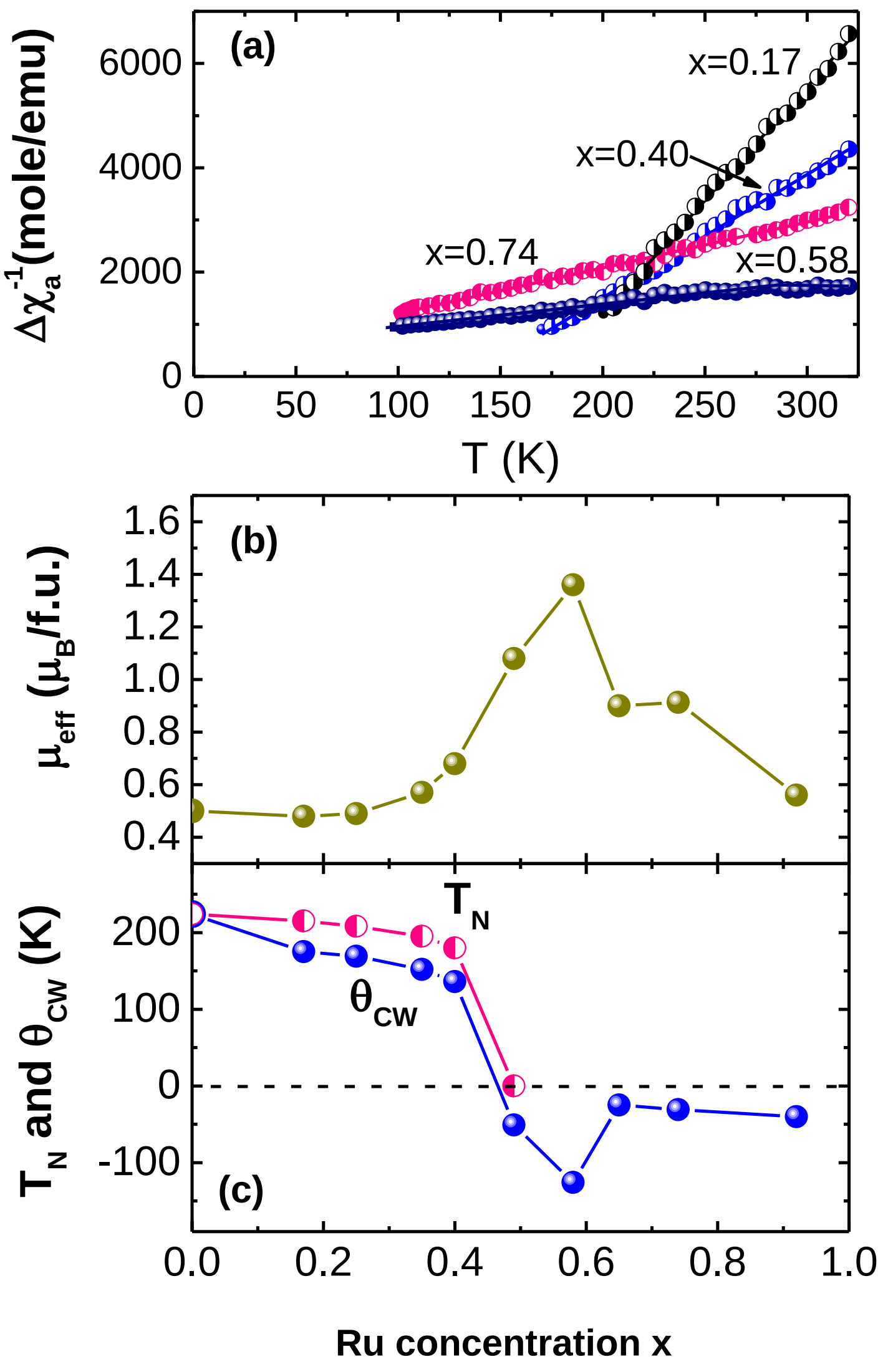

Fig. 5

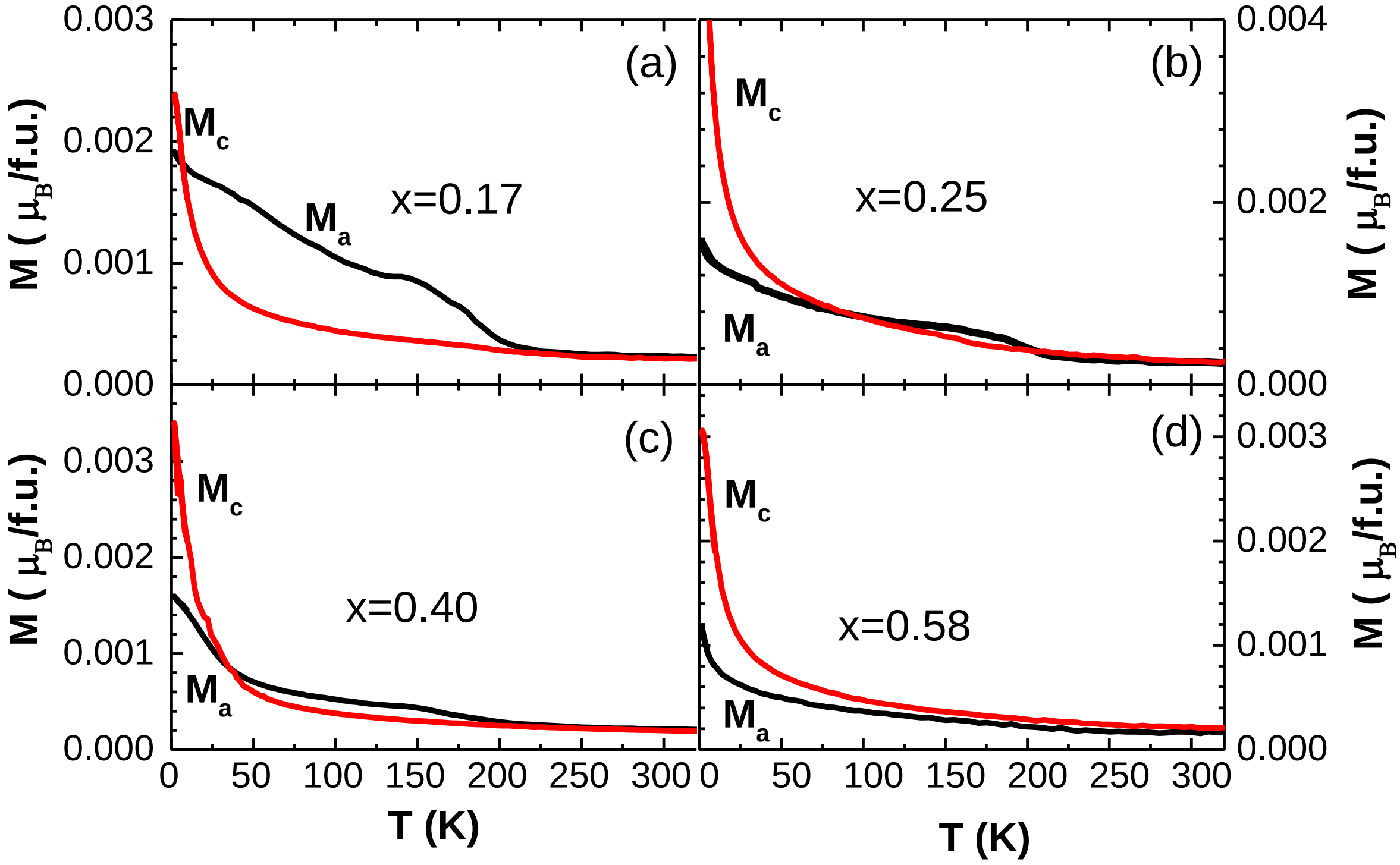

Fig. 6

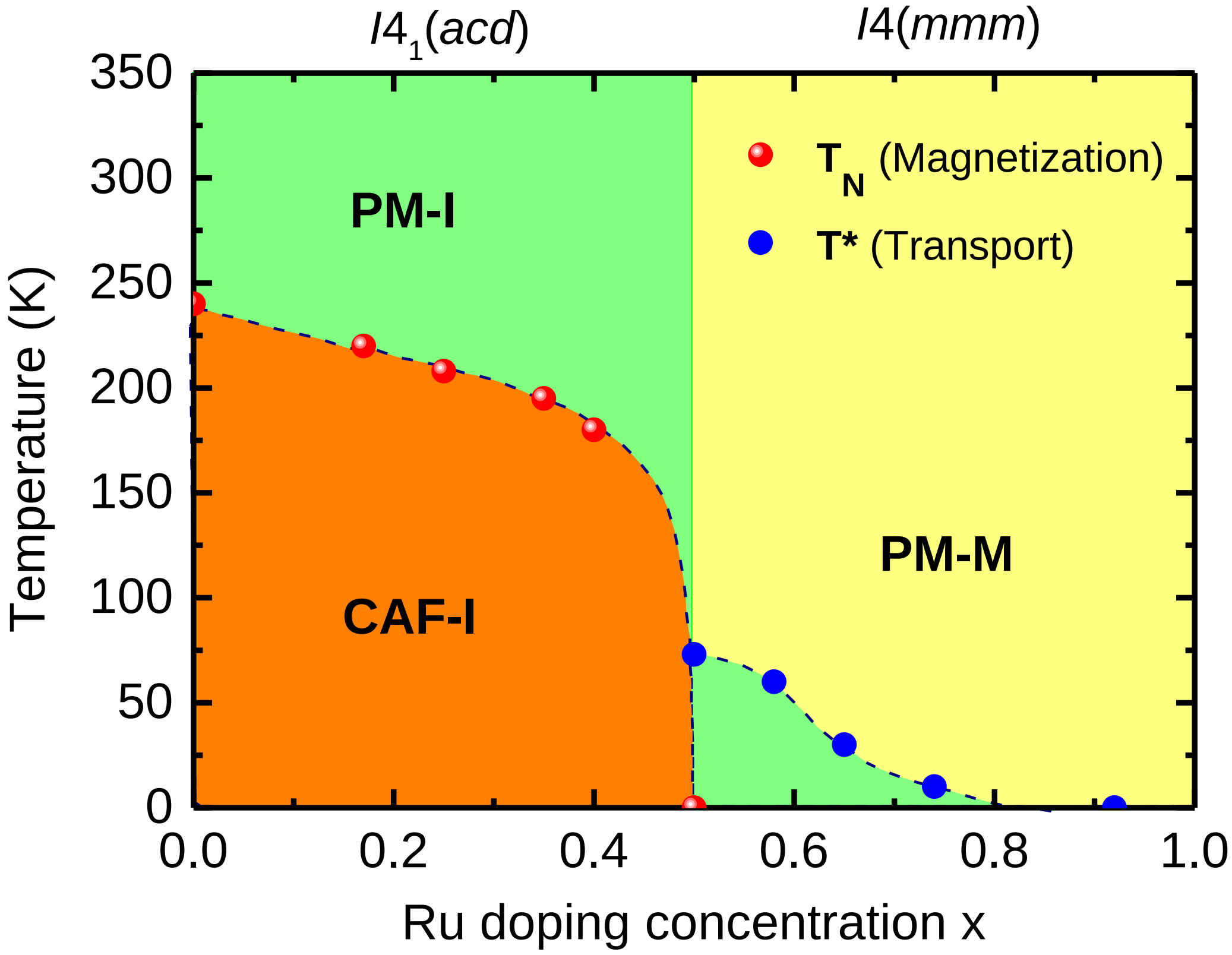

Fig. 7